\documentclass[a4paper,12pt]{article}
\usepackage{graphicx,amsmath,amssymb}  
\usepackage[round]{natbib}
\usepackage{hyperref}
\newcommand{\E}{\mathbb{E}}
\newcommand{\n}{\boldsymbol{n}}
\newcommand{\xib}{\boldsymbol{\xi}}
\newcommand{\phib}{\boldsymbol{\phi}}

\newcommand{\xu}{$\underline{x}$}

\newcommand{\nv}{{\boldsymbol n}}

\newcommand{\nuv}{{\boldsymbol \nu}}

\advance\textheight by \topskip
\begin{document}

\baselineskip 17pt
\parindent 10pt
\parskip 9pt

\begin{titlepage}
\begin{center}
{\Large {\bf  A jump-growth model for predator-prey dynamics: derivation and application to marine ecosystems}}\\ \vspace{1cm} {\large
Samik Datta, Gustav W. Delius, Richard Law}
\\
\vspace{3mm} {\em Departments of Biology and Mathematics,\\ University of York,
\\York YO10 5DD, U.K.\footnote{emails: {\tt sd550, gwd2, rl1 @york.ac.uk} }}
\end{center}

\begin{abstract}
\noindent This paper investigates the dynamics of biomass in a marine ecosystem. A stochastic process is defined in which organisms undergo jumps in body size as they catch and eat smaller organisms. Using a systematic expansion of the master equation, we derive a deterministic equation for the macroscopic dynamics, which we call the deterministic jump-growth equation, and a linear Fokker-Planck equation for the stochastic fluctuations. The McKendrick--von Foerster equation, used in previous studies, is shown to be a first-order approximation, appropriate in equilibrium systems where predators are much larger than their prey.  The model has a power-law steady state consistent with the approximate constancy of mass density in logarithmic intervals of body mass often observed in marine ecosystems.  The behaviours of the stochastic process, the deterministic jump-growth equation and the McKendrick--von Foerster equation are compared using numerical methods.  The numerical analysis shows two classes of attractors: steady states and travelling waves.

\addvspace{0.2 in}

\end{abstract}

\end{titlepage}

\section{Introduction}\label{section.intro}

Marine and freshwater ecosystems exhibit a remarkable regularity in the relation between abundance of organisms and their body masses.  Treating organisms simply as particles of different size, i.e. ignoring taxonomic identity, the total biomass (abundance $\times$ body mass) in logarithmic intervals of body mass is observed to be approximately constant \citep{Sheldon:1972, Sheldon:1977, Boudreau:1992, Kerr:2001}.  Equivalently, the logarithm of abundance expressed as a function of the logarithm of body mass, often referred to as a size spectrum, is approximately linear with a gradient close to $-1$.  Removing the logarithms, this is equivalent to density in mass space being a power function of mass with an exponent $-2$. This approximate regularity applies over a wide range of body size from micro-organisms to large vertebrates, and has been the subject of much research and discussion in ecology \citep{Sheldon:1972, Platt:1978, Heath:1995, Marquet:2005}.

Understanding of the dynamics of biomass flow that lead to this regularity is important: the biomass of most marine ecosystems supports major fisheries that play a significant role in the economies of maritime countries.  The dynamics are often studied by means of a partial differential equation (PDE), in which abundance is taken as a function of both body mass and time. The PDE is motivated by a model of McKendrick \citeyearpar{McKendrick:1926} and von Foerster \citeyearpar{vonFoerster:1959}, in which abundance is a function of age and time.  We will follow the convention of calling this PDE the McKendrick--von Foerster equation, notwithstanding the change in variable from age to size. 

The McKendrick--von Foerster equation was first adopted by Silvert and Platt \citeyearpar{Silvert:1978} in a model allowing growth and mortality to be functions of body mass. Following this, Silvert and Platt \citeyearpar{Silvert:1980} coupled growth at one size to death at another, because organisms grow in size spectra by eating smaller organisms. More recently, the approach has been extended, first to allow organisms to eat those at all smaller sizes \citep{Camacho:2001}, and second, by using a feeding-kernel function, to allow them to eat organisms in a restricted size range \citep{Benoit:2004}. PDEs of this kind are now being used quite extensively to understand processes in marine ecosystems \citep{Andersen:2006, Maury:2007, Andersen:2008}. It can, for instance, be shown in numerical analyses that the PDE at steady state gives size spectra with gradients which are similar to those in marine ecosystems \citep{Blanchard:2008}.

The McKendrick--von Foerster equation is implicitly assumed to be an appropriate approximation for an underlying stochastic process in which individual organisms grow by eating prey items. A first investigation of the relationship between the PDE and the stochastic process \citep{Law:2008} showed that the PDE could describe the approach to a steady-state size spectrum.  However, the stochastic process could also develop travelling-waves; although these were also possible in the PDE, the properties of these waves were somewhat different.  The research described in the present paper was motivated by these discrepancies.

A possible source of these discrepancies is that the McKendrick--von Foerster equation was originally conceived of as a model for organisms indexed by age, rather than by weight. Age and weight do not change in quite the same way over time.  An organism grows older continuously, whereas its weight grows in jumps each time it finds a prey item to feed upon. As time progresses, organisms which start at the same age clearly remain the same age as each other, whereas organisms which start at the same weight in general do not remain the same weight as each other.  Pfister and Stevens \citeyearpar{Pfister:2002} stressed the importance of growth variability in cohorts of organisms.  Motivated by this, Gurney and Veitch \citeyearpar{Gurney:2007} considered the consequences of allowing growth to be both a random variable and also size-dependent, in a von Bertalanffy growth model. However, the emphasis in dynamic size spectra is somewhat different because variation in body weight here emerges from random encounters with prey items of various weights.

In this paper we therefore start from a stochastic process in which organisms undergo jumps in body size as they catch and eat smaller organisms. We introduce this individual-based stochastic process in \ref{subsection.individual} and describe it as a population-level model in Section \ref{subsection.master}. In Section \ref{subsection.sep} we use a systematic expansion of the master equation \citep{vanKampen:1992} to derive an equation for the macroscopic dynamics (which we call the the deterministic jump-growth equation (\ref{macro3})) and a Fokker-Planck equation for the stochastic fluctuations. We also provide an appendix with an alternative derivation of a Langevin equation, following Gillespie \citeyearpar{Gillespie:2000}, to clarify an issue unresolved by the systematic expansion.  Section \ref{subsection.mcKvonF} shows that the McKendrick--von Foerster equation is a first-order approximation of the deterministic jump-growth equation, which applies at steady state when predators are much larger than their prey. In Section \ref{subsection.steadystate} we show that our model has a power-law steady state and we derive an approximate analytic expression for its exponent, thereby showing that the steady state is consistent with the approximate regularity seen in marine ecosystems.  However, the steady state is not necessarily an attractor.  In Section \ref{section.numerics} the behaviour of the deterministic models and of the stochastic model are compared using numerical methods.  As in the case of the McKendrick--von Foerster equation \citep{Law:2008}, certain parts of parameter space allow a travelling-wave solution.
\section{A dynamical model of size-dependent predation}\label{section.maths}

\subsection{An individual-based stochastic process}\label{subsection.individual}

\begin{figure}[ht]
	\centering
		\scalebox{0.8}{\includegraphics{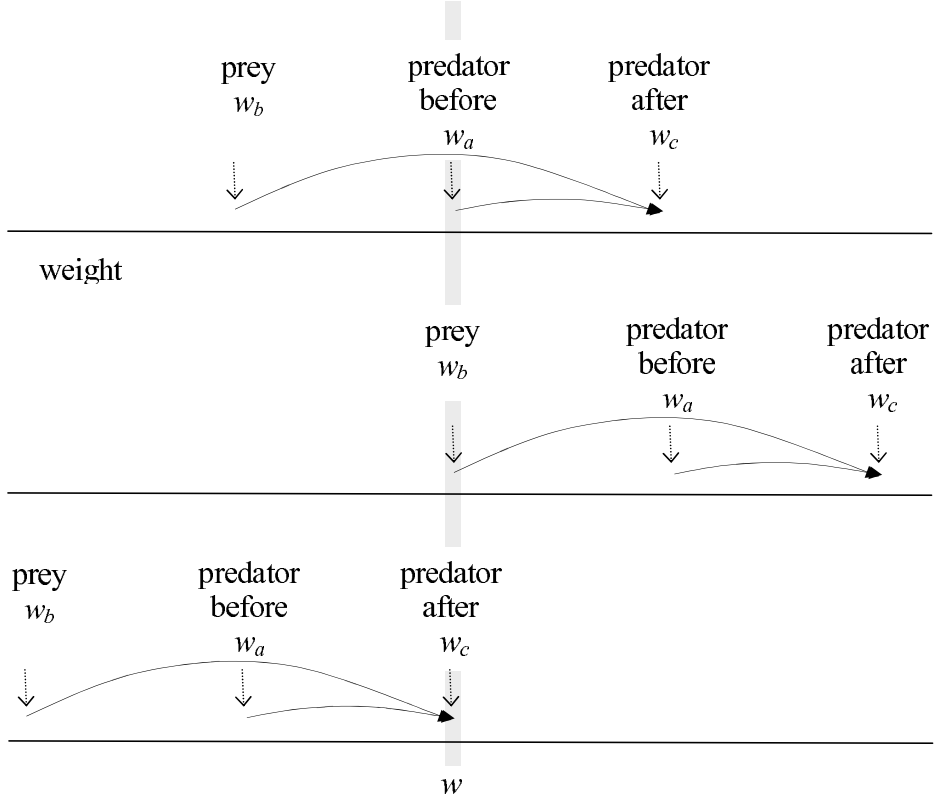}}
 		\caption{
The primary predation event replaces an individual predator and prey by a new, larger predator individual.
Taking some arbitrary weight $w$, there are two ways in which an individual can change from this weight: by feeding and thereby increasing in weight, and by being eaten and so disappearing altogether.  There is also one way in which an individual can become weight $w$: by being of smaller weight and feeding on a prey of just the right size to become weight $w$. These events are reflected in the three terms of the deterministic jump-growth equation (\ref{macro3}) in Subsection \ref{subsection.djge}.}
  \label{fig.events}
\end{figure}

We model predation as a Markov process. The primary stochastic event comprises a predator of weight $w_a$ consuming a prey of weight $w_b$ and, as a result, increasing to become weight $w_c$ (Figure \ref{fig.events}).  Predation is inefficient and, in keeping with ecological convention, we assume that a fixed proportion $K$ of prey mass is assimilated by the predator so that $w_c = w_a+Kw_b$ (the assumption of constant $K$ could be relaxed). We call this model the 'jump-growth model' because the changes in the weight distribution are caused by discrete steps in body size as predators eat prey, and the mortality that comes with this predation.

It would be easy to add additional events to the jump-growth model to account for natural death and for birth (recruitment) but, as we will see, for the purpose of this paper of explaining the observed power law size spectrum these additional events are not required, and we will therefore restrict our attention to the pure predation events.

The next three subsections will be concerned with the derivation of equations describing the time evolution of the weight distribution that follows from this stochastic process. The main result from these sections that we will use further in this paper is the deterministic jump-growth equation \eqref{macro3} given in Section \ref{subsection.djge}. That equation has an intuitive explanation in figure \ref{fig.events} and is enough to follow the remainder of the paper.

A mathematically rigorous treatment of the individual-based model may be possible following the techniques developed for stochastic processes on configuration-space, see for example \citep{Finkelshtein:2009}. In this paper we will instead pursue a heuristic treatment of a corresponding population-level model.

\subsection{A population-level master equation}\label{subsection.master}

Instead of keeping track of the weight of each individual, we aggregate individuals of similar weights into weight brackets, and follow the number of individuals in each bracket. We introduce a set of weights $w_i$ and corresponding weight brackets  $[w_i,w_{i+1})$, with $i\in \mathbb{Z}$. In practice, the size of the weight brackets should be chosen small enough so that discretisation errors are small. The weight distribution of organisms in a large fixed volume $\Omega$ is described by a sequence of numbers "$[\dots,n_{-1},n_0,n_1,\dots]$, where $n_i$ is the number of organisms in $\Omega$ with weights in the $i$-th weight bracket between $w_i$ and $w_{i+1}$.

Let $k_{ij}/\Omega$ denote the rate constants for the predation events, where the indices of $k$ are ordered: predator before feeding, prey.  Then the probability in an infinitesimal time interval $dt$ for any one of $n_i$ organisms in weight bracket $i$ to eat any one of the $n_j$ organisms of weight bracket $j$ is $k_{ij}\Omega^{-1}n_in_jdt$. The time evolution of the probability $P(\n,t)$ that the system is in the state $\n$ at time $t$ is then given by the master equation
\begin{equation}\label{meqn2}
 \frac{\partial P(\n,t)}{\partial t} = \sum_{i,j} \frac{k_{ij}}{\Omega}\left[(n_i+1)(n_j+1)P(\n-\boldsymbol{\nu}_{ij},t)-n_i n_j P(\n,t)\right],
\end{equation}
where $\n-\boldsymbol{\nu}_{ij} = (\dots,n_j+1,\dots,n_i+1,\dots,n_l-1,\dots)$, and $l$ is the index of the weight bracket $w_l\leq w_i+Kw_j<w_{l+1}$. The first (positive) term in (\ref{meqn2}) corresponds to having one extra predator in weight bracket ($i$), one extra prey in ($j$), and one less predator in ($l$), so that one predation event will move the system from state $\n-\boldsymbol{\nu}_{ij}$ into state $\n$.  The second (negative) term corresponds to another such predation event that moves the system out of state $\n$. Hence the master equation is commonly referred to as a ``gain-loss'' equation.

\subsection{Separation of macroscopic behaviour and fluctuations}\label{subsection.sep}

The master equation \eqref{meqn2} has non-linear coefficients and is difficult to solve analytically. We therefore make use of the property that, in systems of sufficiently large volume $\Omega$, the fluctuations are relatively small because they are suppressed by a factor of the square root of $\Omega$; the conditions required for this to be true are given in Appendix \ref{gillespie}.  In this section we adopt the approach of van Kampen \citeyearpar{vanKampen:1992}, carrying out an expansion of \eqref{meqn2} in $\Omega$, and collecting together the highest-order terms in $\Omega$.  To do this, it helps to rewrite the master equation \eqref{meqn2} using a step-operator notation:
\begin{equation}\label{meqn}
 \frac{\partial P(\n,t)}{\partial t} = \sum_{i,j} \frac{k_{ij}}{\Omega}\left(\E_i\E_j\E_l^{-1}-\mathbb{I}\right)\left(n_i n_j P(\n,t)\right).
\end{equation}
Here $\E_i$ is a step operator that acts on a function $f(\n)$ to give $\E_i f([\dots,n_i,\dots])$ = $f([\dots,n_i+1,\dots])$;  similarly $\E_j$ acts on a function $f(\n)$ to give $\E_j f([\dots,n_j,\dots])$ = $f([\dots,n_j+1,\dots])$;  conversely $\E_l^{-1}$ acts on a function $f(\n)$ to give $\E_l^{-1} f([,\dots,n_l,\dots])$ = $f([\dots,n_l-1,\dots])$.  Thus \eqref{meqn} is just an alternative notation for \eqref{meqn2}.  For further explanation of the step-operator notation, see van Kampen \citeyearpar[page 139]{vanKampen:1992}.

Following the method used by van Kampen \citeyearpar{vanKampen:1992}, we separate each random variable $n_i$ into a deterministic component $\phi_i(t)$ which describes the density of individuals in weight bracket $i$, and a random fluctuation component $\xi_i(t)$ as
\begin{equation}\label{ansatz}
 n_i = \Omega\phi_i(t) + \Omega^{\frac{1}{2}} \xi_i(t).
\end{equation}
On average the number of individuals will be proportional to the system size $\Omega$, by the law of large numbers, and that is the reason for the factor of $\Omega$ multiplying $\phi_i(t)$. That the fluctuations are proportional to the square root of the system size should be justified by some sort of central limit theorem. A heuristic justification is given in appendix \ref{gillespie}.
Thus disaggregating $n_i$ in this way leaves two variables $\phi_i$ and $\xi_i$ which no longer scale with the system size. We assume that $\Omega$ is so large that the discrete nature of $\n$ is no longer noticeable at the level of $\phib$ and $\xib$ and we can treat them as continuous variables.

The new random variables $\xi_i$ are described by a probability distribution $\Pi(\xib,t)=\Omega^{1/2}P(\n,t)$. An equation for the time evolution of this probability distribution is obtained by substituting the change of variables \eqref{ansatz} into the master equation \eqref{meqn}. Care needs to be taken because this change of variables is time-dependent. This has the consequence that
\begin{equation}
  \frac{\partial P(\n,t)}{\partial t} = \Omega^{-1/2}\frac{\partial \Pi(\xib,t)}{\partial t}-
  \sum_i\frac{d\phi_i}{dt}\frac{\partial\Pi(\xib,t)}{\partial\xi_i}.
\label{eq:pt}
\end{equation}
Here we used the property that $\Omega^{-1/2}d\xib/dt =  -d\phib/dt$ when we keep $\n$ fixed.
The operators $\E_i$ which change $n_i$ to $n_i+1$ now change $\xi_i$ to $\xi_i+\Omega^{-1/2}$ and can therefore be expanded as
\begin{equation}
\E_i=1+\Omega^{-1/2}\frac{\partial}{\partial\xi_i}+\frac12\Omega^{-1}\frac{\partial^2}{\partial\xi_i^2}+\cdots.
\label{eq:eexp}
\end{equation}
Substituting all this into the master equation \eqref{meqn} gives an equation with terms containing various different powers of the system size $\Omega$.

The highest order terms are at order $\Omega^{0}$.  They only contain the macroscopic variables $\phi_i$ and vanish if these satisfy the deterministic equation
\begin{equation}\label{macro2}
 \frac{d\phi_i}{dt}= \sum_{j} 
 \left( -k_{ij} \phi_i\phi_j -k_{ji} \phi_j\phi_i+ k_{mj} \phi_m\phi_j \right),
\end{equation}
where $m$ is an index for the weight bracket: $w_m\leq w_i-Kw_j<w_{m+1}$. The three terms in \eqref{macro2} are in keeping with the intuition given by Figure \ref{fig.events}.  Losses from weight bracket $i$ (the negative terms) occur because individuals in this bracket eat prey and become heavier, and because these individuals are themselves eaten. Gains into weight bracket $i$ (the positive term) occur through smaller predators growing into this bracket by eating prey. Imposing the deterministic equation \eqref{macro2} is not the only possible way to make the terms of order $\Omega^{0}$ vanish, but it is the most natural and is independently derived in appendix \ref{gillespie}.

Terms at the next order, $\Omega^{-1/2}$, give the linear Fokker-Planck equation for the probability distribution $\Pi(\xib)$ of the fluctuations,
\begin{equation}
\frac{\partial \Pi}{\partial t}=
-\sum_{ij}A_{ij}\frac{\partial}{\partial\xi_i}\left(\xi_j\Pi\right)
+\frac12\sum_{ij} B_{ij}\frac{\partial^2}{\partial\xi_i\partial\xi_j}\Pi,
\end{equation}
where the coefficients $A_{ij}$ and $B_{ij}$ are independent of the fluctuations $\xib$. If we introduce the objects $k_{ijl}$ and $f_{ijk}$ by
\begin{align}
k_{ijl} &= \left\{\begin{array}{ll}k_{ij}&\text{ if } w_l\leq w_i+Kw_j < w_{l+1}\\ 0 &\text{ otherwise}\end{array}\right.,\\
f_{ijl} &= \frac{1}{2}\left(k_{ijl}+k_{jil}\right)
\end{align}
then we can give the succinct expressions
\begin{align}
A_{ii}&=\sum_{jl}f_{ijl}\phi_j,&
A_{ij}&=\sum_l\left(f_{ijl}\phi_i-f_{lji}\phi_l\right),\\
B_{ii}&=\sum_{jl}f_{jli}\phi_j\phi_l,&
B_{ij}&=\sum_l\left(f_{ijl}\phi_i\phi_j-f_{ilj}\phi_i\phi_l-f_{lji}\phi_l\phi_j\right).
\end{align}

Because the fluctuations are damped by a factor of $\Omega^{1/2}$, in the remainder of this paper we concentrate on studying the deterministic equation \eqref{macro2}.

\subsection{The deterministic jump-growth equation}\label{subsection.djge}

For analytical calculations and also for conceptual considerations it is convenient to work with the continuum limit of the macroscopic equations \eqref{macro2}. This limit is obtained by writing the size of the weight brackets as $\Delta_i = w_{i+1}-w_i = \mu_i \Delta$ and taking the limit $\Delta\rightarrow0$. The discrete set of variables $\phi_i$ is replaced by a continuous density function $\phi(w)$ satisfying $\phi(w_i)=\phi_i/\Delta_i$. This function $\phi(w)$ describes the density per unit mass per unit volume as a function of mass $w$ at time $t$; it therefore has dimensions M$^{-1}$ L$^{-3}$. The sum over weights in \eqref{macro2} is replaced by an integral, $\sum_i\Delta_i \rightarrow \int dw$. The rate constants $k_{ij}$ are replaced by a feeding rate $k(w,w')$ satisfying $k(w_i,w_j)=k_{ij}$.  The macroscopic equation \eqref{macro2} becomes
\begin{align}\label{macro3}
  \frac{\partial\phi(w)}{\partial t} = \int
    (&-k(w,w')\phi(w)\phi(w') \nonumber\\ 
     &-k(w',w)\phi(w')\phi(w) \nonumber\\
          &+k(w-Kw',w')\phi(w-Kw')\phi(w'))dw'.
\end{align}
We call this equation the 'deterministic jump-growth' equation.
The three terms in (\ref{macro3}) are equivalent to those in (\ref{macro2}), and correspond to the idea in Figure (\ref{fig.events}) that there are two ways to leave weight $w$ and one way to enter it. The terms represent, in order: feeding on prey to become larger than weight $w$, being fed upon and removed from the system entirely, and feeding on prey of exactly the right size to become weight $w$.

Following  Beno\^it and Rochet \citeyearpar{Benoit:2004} we assume that the feeding rate takes the form
\begin{equation}\label{feedrate}
  k(w,w')=Aw^{\alpha}s\left(w/w'\right).
\end{equation}
This states that the rate at which a particular predator of weight $w$ eats a particular prey of weight $w'$ is a product of the volume searched per unit time and a dimensionless feeding preference function $s$. The volume searched per unit time $Aw^{\alpha}$ depends on the predator's body weight $w$, raised to the constant power $\alpha$. $A$ is a constant volume searched per unit time per unit $\mathrm{mass}^{\alpha}$. The feeding preference function $s$ depends only on the ratio $w/w'$ between predator weight and prey weight. In practice this feeding preference function will be peaked around a preferred predator:prey weight ratio $B$.

When the parameter $K$, that describes which proportion of the prey mass is assimilated by the predator, is equal to 1, the deterministic jump-growth equation \eqref{macro3} reduces to the Smoluchowski coagulation equation \citep{Smoluchowski:1916}, that is used to describe the clumping together of particles, for example in aerosols. However the rate kernels used to describe coagulation differ greatly from our localised feeding rate kernel \eqref{feedrate}. Typical choices in the coagulation equation are $k(x,y)=x+y$ or $xy$ or other homogeneous expressions and these lead to very different behaviour such as the formation of one giant cluster (gelation); see for example \cite{Aldous:1999}.

\subsection{Relation to the McKendrick--von Foerster equation}\label{subsection.mcKvonF}

The deterministic jump growth equation \eqref{macro3} is not the same as the McKendrick--von Foerster equation which has been widely used to describe the dynamics of size spectra \citep{Silvert:1978, Silvert:1980, Benoit:2004, Maury:2007, Blanchard:2008, Law:2008} and which reads
\begin{equation}
\frac{\partial \phi}{\partial t}= - \phi D -\frac{\partial}{\partial w}(\phi G),
\label{mvf}
\end{equation}
where $D$ is the per capita death rate at weight $w$ from predation by larger organisms,
\begin{equation}
 D(w) = \int k(w',w) \phi(w') dw',
\label{mvfD}
\end{equation}
and $G$ is the growth rate at weight $w$ from feeding on smaller organisms, 
\begin{equation}
 G(w) = \int Kw' k(w,w') \phi(w') dw'.
\label{mvfG}
\end{equation}

Here we show that \eqref{mvf} emerges as an approximation to \eqref{macro3} in the case where the typical prey is small in size compared with the predator. Such an assumption is reasonable in many cases, because predators tend to be of an order $10^2$ to $10^3$ times the body mass of their prey \citep{Cohen:1993, Jennings:2003}.  So the feeding kernel is strongly peaked around $w'=w/B$ with $B$ large. Taking into account further the inefficiency with which prey mass is assimilated ($K\approx10^{-1}$) \citep{Paloheimo:1966}, there is some justification for treating $Kw'<<w$ in the last term of \eqref{macro3}. This motivates a Taylor expansion of this term around $w$,
\begin{align}
k(w-Kw',w')\phi(w-Kw') = &\ k(w,w')\phi(w) \nonumber\\
&+(-Kw')\frac{\partial}{\partial w}\left(k(w,w')\phi(w)\right) \\
&+\frac{(-Kw')^2}{2!}\,\frac{\partial^2}{\partial w^2}\left(k(w,w')\phi(w)\right)
+\cdots\nonumber
\end{align}
Substituting this back into \eqref{macro3} gives
\begin{align}\label{macro5}
\frac{\partial\phi(w)}{\partial t} = 
& -\int k(w',w)\phi(w)\phi(w') dw' \nonumber \\
& - \frac{\partial}{\partial w}\
\int Kw' k(w,w')\phi(w)\phi(w') dw' \\
& + \frac{1}{2}\frac{\partial^2}{\partial w^2}\
\int (Kw')^2 k(w,w')\phi(w)\phi(w') dw' \nonumber \\
& +R, \nonumber
\end{align}
where the remainder term R is given by
\begin{equation}\label{fluccor}
R=\sum_{n=3}^\infty  \frac{(-1)^n}{n!}\frac{\partial^n}{\partial w^n} \int (Kw')^n k(w,w')\phi(w)\phi(w')dw'.
\end{equation}
The first two terms in \eqref{macro5} correspond to those in the McKendrick--von Foerster equation \eqref{mvf}. For ecosystems near to steady state, where $\phi(w)$ is close to a power law (as we will see in the next section), the higher order terms are suppressed by factors of $K/B$ and are therefore small. Thus the McKendrick--von Foerster equation is a good approximation for \eqref{macro3} near the steady state and when prey are typically much smaller than their predators.  But the higher order terms are not necessarily small in non-equilibrium ecosystems.  In particular, the McKendrick--von Foerster equation is a less good approximation if there is a travelling wave attractor, see Section \ref{subsection.waves}.

One way to understand the difference between \eqref{macro3} and \eqref{mvf} is that \eqref{macro3} models the discrete, variously-sized jumps in predator mass as predators feed and grow.  This captures the property of the stochastic model that individuals, starting at a given weight, are able to develop a range of weights over the course of time.  In contrast to this, the McKendrick--von Foerster equation \eqref{mvf} assumes smooth growth along the weight axis.  Spreading of body size can be incorporated in \eqref{mvf} by introducing the diffusion term, the third term in \eqref{macro5}.  The source of such diffusion is the deterministic jump-growth equation (i.e. terms of order $\Omega^0$), so diffusion is attributable to the deterministic, as opposed to the stochastic, component of the full process.

\subsection{Steady-state solution}\label{subsection.steadystate}

In marine ecosystems, abundance of organisms within body mass classes averaged over space and seasons often changes rather little, suggesting that they may be close to a steady state. In such circumstances and when abundance and mass are both expressed as logarithms (i.e. as a power spectrum) the relationship is approximately linear with a gradient often close to -1, which implies a power law with an exponent -2 in the untransformed variables.  This leads to the important regularity of marine ecosystems that the total biomass is approximately constant when expressed in logarithmic intervals of body mass.   

Beno\^it and Rochet \citeyearpar{Benoit:2004} found that the McKendrick--von Foerster equation has steady state solutions of the power law form $\hat{\phi}(w)\propto w^{-\gamma}$ \citep[see also][]{Platt:1978, Camacho:2001}, and we will show that the same is true for the deterministic jump-growth equation \eqref{macro3}. Of course in the real world such a power law will have to break down for very small weights (where otherwise the power law would predict an infinite number of very small individuals) and for very large weights (where the power law would predict a non-zero density of arbitrarily large individuals). Indeed, in a real system with a finite number of individuals, a model just having predation events could not have a non-trivial steady state because the number of individuals would continue to decrease. A non-zero steady state is possible only if there is an inexhaustible reservoir of small individuals. In our model the power law spectrum provides this reservoir automatically. In a more realistic model one would need to model the plankton as well as recruitment.

A steady state solution $\hat{\phi}(w)$ of (\ref{macro3}) has to satisfy the equation
\begin{eqnarray}\label{steadystate}
0 = &-&    \int k(w,w')\hat{\phi}(w)\hat{\phi}(w')  dw' \nonumber \\
    &-& {} \int k(w',w)\hat{\phi}(w')\hat{\phi}(w)  dw' \\
    &+& {} \int k(w-Kw',w')\hat{\phi}(w-Kw')\hat{\phi}(w')  dw',\nonumber 
\end{eqnarray}
If we substitute the power law Ansatz $\hat{\phi}(w)\propto w^{-\gamma}$ into this equation, use the form \eqref{feedrate} for the feeding rate, change to a new integration variable $r=w_{predator}/w_{prey}$ and cancel some overall factors, we obtain an equation for the exponent $\gamma$,
\begin{equation}\label{steadystate1}
0 = f(\gamma) = \int s(r) \Bigg(-r^{\gamma-2}-r^{\alpha-\gamma}+r^{\alpha-\gamma}(r+K)^{-\alpha+2\gamma-2}\Bigg)\ dr.
\end{equation}
The existence of a power law steady state can now be proven using the same argument as that given by Beno\^it and Rochet \citeyearpar{Benoit:2004} in the case of the McKendrick--von Foerster equation\footnote{We thank one of the referees for pointing this out.}. The argument goes as follows. If we assume that predators are bigger than their prey, then for $\gamma<1+\alpha/2$, $f(\gamma)$ is less than zero.  Also, $f(\gamma)$ increases monotonically for $\gamma>1+\alpha/2$, and is positive for large positive $\gamma$. Therefore there will always be a unique $\gamma$ for which $f(\gamma)$ is zero and thus a unique steady state of the form $\hat{\phi}(w)\propto w^{-\gamma}$. If we allow predators to be smaller than their prey, situations with no power law steady state or multiple power law steady states can be found.

The numerical value of the power law exponent $\gamma$ is of particular interest because $\gamma$ is known to have a value close to 2 in marine ecosystems (see Section \ref{section.intro}).  In the special case that $K=1$, and $\alpha=1$, a value $\gamma=2$ does in fact satisfy \eqref{steadystate1}.  A value of $\alpha$ close to 1 is biologically reasonable as this means that the volume searched by a predator is proportional to its body weight (see Equation \eqref{feedrate}), although the limited information available suggests a value slightly lower than $\alpha=1$ \citep{Ware:1978}.  More generally, $\gamma=(3+\alpha)/2$ will satisfy \eqref{steadystate1} for any $\alpha$, with $K=1$.

A value of $K$ close to 1 is unrealistic: $K\approx0.1$ would be more appropriate \citep{Paloheimo:1966} because only a small proportion of food ingested is assimilated into extra body weight.  To treat this case analytically we make the assumption that predators feed only on prey of their preferred size, i.e., we set the feeding preference function in \eqref{steadystate1} to the delta function $s(r)=\delta(r-B)$. In that case \eqref{steadystate1} reduces to 
\begin{equation}\label{steadystate2}
  0=-B^{\gamma-2}-B^{\alpha-\gamma}+B^{\alpha-\gamma}(B+K)^{-\alpha+2\gamma-2}.
\end{equation}
A Taylor expansion in $K/B$ gives
\begin{equation}\label{steadystate3}
  0 \approx(2\gamma-\alpha-2)\frac{K}{B} - B^{-2\gamma+\alpha+2},
\end{equation}
and the Lambert $W$ function can be used to express $\gamma$ explicitly as a function of the other variables
\begin{equation} \label{gamma}
  \gamma \approx \frac{1}{2} \left( 2 + \alpha +
   \frac{W\left(\frac{B}{K}\log{B}\right)}{\log{B}}\right).
\end{equation}

At $K=1$ and $\alpha=1$, \eqref{gamma} produces $\gamma = 2$ because $W(B\log B)=\log B$.  For $K<1$, the exponent $\gamma$ increases as either $K$ or $B$ decrease, because in either case less mass is transferred to larger organisms.  Notice however that the dependence of $\gamma$ on $K$ and $B$ is weak; for instance, if $K=0.1$ and $B=100$ (still with $\alpha=1$), the exponent only increases to $\gamma=2.21$.  Thus if $K$ and $B$ are given biologically reasonable values the steady-state of the model is broadly consistent with the empirical property of marine ecosystems that $\gamma$ is close to 2.  

The ecological literature contains a relationship between the parameter $\gamma$, and $K$ and $B$ based on a quite different premise, that the metabolic rate of organisms scales with body weight as $w^{3/4}$.  It can be shown from this scaling that
\begin{eqnarray} \label{brown}
  \gamma = 1+\frac{3}{4} - \frac{\log{K}}{\log{B}}
\end{eqnarray}
in the absence of any consideration of dynamics \citep{Brown:2004}.  There is some resemblance between this equation and \eqref{gamma}, which becomes evident from taking the asymptotic approximation for the Lambert $W$ function
\begin{equation}
  W(z)=\log{z}-\log{\log{z}}+\cdots
\end{equation}
in \eqref{gamma}, giving an expansion in which the leading terms are
\begin{equation} \label{gammaa}
  \gamma \approx \frac{1}{2} \left( 3 + \alpha - \frac{\log{K}}{\log{B}}
  -\frac{\log{\log\left(\frac{B}{K}\log B\right)}}{\log B} 
  +\frac{\log{\log{B}}}{\log{B}}+\cdots\right).
\end{equation}
Both (\ref{brown}) and (\ref{gammaa}) contain the term ($\log K$)/($\log B$), but are not the same. From a biological standpoint the equations have the important difference that the relationship in (\ref{gammaa}) follows simply from dynamical bookkeeping of biomass, without any assumption about metabolic rates being made (see also Law et al. \citeyearpar{Law:2008}).

We stress that, although some properties of the steady state have been described here, we have not investigated analytically the region of parameter space in which the steady state is an attractor.  The next Section (\ref{section.numerics}) shows by means of numerical methods two classes of attractor: a steady state of the kind described above and a travelling wave.

\section{Numerical results}\label{section.numerics}

Here we use numerical methods to compare some properties of the stochastic jump-growth model  \eqref{meqn}, the deterministic jump-growth equation \eqref{macro3} and the McKendrick--von Foerster equation \eqref{mvf}. 

Body sizes can span at least ten orders of magnitude in real ecosystems, and it is helpful in computations to discretise weight into logarithmic bins, such that the weight bracket $[w_i,w_{i+1})$ is the range $[w_i,(1+\Delta)w_i)$.  We adopt a notation: $x = \log(w/w_0)$, for some arbitrary weight $w_0$, and use the function $u(x) = \Omega w \phi(w)$.  Thus, integrating $u(x)$ over the range $[x_i,x_i+\Delta)$, returns the total number of individuals in this size range.

Some further biological details have to be specified to do the numerical analysis;  Table \ref{table.parameters} summarises the information, and Section \ref{subsection.numerics} describes this in more detail. We have chosen the parameters not for biological realism but in order to highlight the differences between the stochastic jump-growth model and the McKendrick--von Foerster equation. In particular we have chosen a smaller predator:prey mass ratio than is realistic.

\begin{table}[ht]
  \begin{center}
    \begin{tabular}{cclll}
      term     & meaning                         & & value &\\
               &      & Fig \ref{fig.timeseries} & Fig \ref{fig.growth}& Fig \ref{fig.stability} \\
      \hline
      \xu 		 & min wt of phytoplankton         &0           &0           &0           \\
      $x_b$    & min wt of consumers             &2           &2           &2           \\
      $x_d$    & max wt of newborn consumers     &2.1         &2.1         &2.1         \\
      $x_s$    & wt at start of senescent death  &5           &7           &8           \\
      $\overline{x}$  & max wt of consumers      &7.5         &9           &10          \\
      $K$      & mass conversion efficiency      &0.2         &0.2         &0.2         \\
      $B$      & preferred pred:prey mass ratio &$e^1$       &$e^1$       &$e^1$       \\
      $A$      & volume searched mass$^{-\alpha}$&50          &50          &50          \\
      $\alpha$ & search volume exponent          &1           &1           &1           \\
      $\sigma$ & width of feeding kernel         &0.3         &0.35        &variable    \\
      $\mu$    & intrinsic mortality rate        &0.1         &0.1         &0.1         \\
      $\rho$   & growth of senescent death       &5           &5           &5           \\
      \hline
      &\textit{stochastic realisation}\\
      $N_p$    & number of phytoplankton         &25000       &50000       &-       \\
      $N_0$    & initial number of consumers     &2000        &4000        &-        \\
      $x_0$    & initial upper bd of consumers   &4           &7           &-           \\
      $\gamma^*-1$&exponent for fixed spectra    &1.3         &1.3         &-         \\
      $\Delta'$& weight bracket for stochastic bins&0.1       &0.1         &-           \\
      \hline
      &\textit{numerical integration}\\
      $\Delta$ & wt bracket for integration      &0.01        &0.01        &0.01        \\
      $\delta t$& time increment for integration &0.0001      &0.0001      &0.0001      \\
    \end{tabular}
  \caption{\small{Parameter meanings and values used in computations for figures}}
  \label{table.parameters}
  \end{center}
\end{table}

\subsection{Model specification for numerics}\label{subsection.numerics}

The numerical results describe an ecosystem with two types of organism: phytoplankton which do not feed on other organisms, and consumers which feed on each other and on phytoplankton.  In more detail, the full range of body weights $[\underline{x},\overline{x})$ is subdivided into the following regions with different ecological properties.
\begin{itemize}
\item $[\underline{x},x_b)$ is reserved for phytoplankton.  These organisms are self-supporting; they do not change in mass, and do not form part of the dynamics.  Their densities are held constant, which is equivalent to assuming that, as soon as they are eaten, they are replaced.  Such organisms have to be present to provide a supply of food for small consumers.
\item $[x_b,x_d)$ is a range reserved for renewal of consumers, i.e. a range over which consumers are born.  Renewal is essential: without this, consumers would gradually die out.  Biological realism requires this range to be distinguished from $[\underline{x},x_b)$, because newborn consumers may grow in size.  When consumers leave this range (by growth or by death), they are immediately replaced, which amounts to an assumption of perfect density-dependent compensation in the nursery.
\item $[x_d,x_s)$ is the range in which consumers experience the standard predation, growth and death processes described in Section \ref{section.maths}.  We include in this range intrinsic mortality at a per-capita rate $\mu$, which takes into account the fact that organisms can die for reasons other than being eaten.
\item $[x_s,\overline{x})$ is a range in which the per-capita mortality rate of consumers increases according to the function
\begin{equation}
  d(x) = 
  \left\{
  \begin{array}{ll}\label{mortality}
    \mu\exp\left(\rho(x-x_s)\right)  & \text{if}\ x \geq x_s          \\
    \mu                                   & \mbox{otherwise}
  \end{array}
  \right.
\end{equation}
where $\rho$ scales how fast mortality increases beyond size $x_s$.  The purpose of this is to ensure that consumers cannot continue to grow indefinitely, in keeping with biological constraints on body size.  The upper bound $\overline{x}$ is set such that the density of organisms at this size is very close to zero.
\end{itemize}  

For numerical studies, the predation-rate function $k(x,x')$ needs to be made explicit.  In keeping with \eqref{feedrate}, this function is taken to consist of a volume searched per unit time by predators, together with a feeding preference function, which is assumed to have a Gaussian shape.  In logarithmic variables, the function is:
\begin{equation}
  k(x,x') = 
  \left\{
  \begin{array}{ll}\label{kernel}
    \frac{Ae^{\alpha x}}{\sigma \sqrt{2\pi}}\exp\left( -\frac{1}{2\sigma^2} (x-x'-\log B)^2 \right)  & \text{if}\ x > x'          \\
    0                                   & \mbox{otherwise}
  \end{array}
  \right.
\end{equation}
where parameters $A,\alpha,B$ remain as defined in Section \ref{subsection.steadystate}, and $\sigma$ measures the range of prey sizes likely to be eaten relative to the size of the predator. We have introduced the assumption here that predators must be larger than their prey.

 In stochastic realisations, the fixed phytoplankton population was initialised with $N_p$ individuals taken from an exponential distribution with parameter $\gamma^*-1$ over the range $[\underline{x},x_b)$.  The consumer spectrum was initialised with $N_0$ individuals taken from an exponential distribution with parameter $\gamma^*-1$ over a range $[x_b,x_0)$.  $N_0$ was chosen to make the discontinuity between the two spectra small,  the upper weight limit being initially $x_0$ in the consumers.  After the start, consumers dying or growing out of the renewal range were replaced with newborn individuals, using the same exponential distribution so that the number of consumers in this range would remain constant.  We carried out realisations of the individual-based stochastic process (Subsection \ref{subsection.individual}) using the Gillespie algorithm \citep{Gillespie:1976}. Body sizes were aggregated into bins of width $\Delta'$ only for visualisation of the size spectra.
 
Numerical integrations of the deterministic models were carried out using the explicit Euler method, with a bin width $\Delta$ and a time step $\delta t$;  consumer spectra were held at their initial values in the renewal range.  Integrations were initialised with assumptions equivalent to those of the corresponding stochastic realisations.  For graphical comparison with stochastic results, $u(x)$ was scaled such that $\int{u(x,0)dx}$ was $N_p$ and $N_0$ for the phytoplankton and consumers respectively, and displayed as the number $n(x) =u(x)\Delta'$ over size intervals $\Delta'$.

\subsection{Travelling waves\label{subsection.waves}}

\begin{figure}[ht]
	\centering
  	\includegraphics[width=5in, height=1.5in]{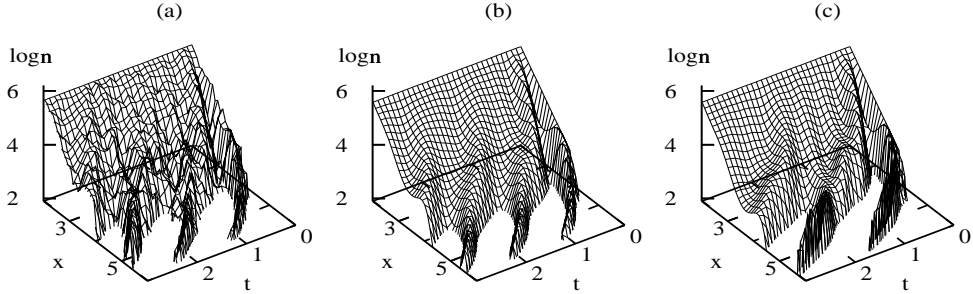}
		\caption{\small{Size spectra expressed as logarithm of numbers $\log n(x)$ with logarithm of weights $x$ over time $t$, constructed from (a) the stochastic jump-growth process, (b) the deterministic jump-growth equation, (c) the McKendrick--von Foerster equation.  Parameter values given in Table \ref{table.parameters}.}}
  \label{fig.timeseries}
\end{figure}

Figure \ref{fig.timeseries} compares time series from the deterministic jump-growth equation \eqref{macro3} and from the McKendrick--von Foerster equation \eqref{mvf} against a realisation of the stochastic process.  Parameter values are the same for all three time series, and were chosen to contrast the two deterministic models, by making the difference between predator and prey body sizes relatively small, and by ensuring the steady state would not be an attractor.  Initial conditions were chosen well away from the steady state, to induce large oscillations in the size spectra from the start.

Large sustained waves in density develop over time in all three models.  These waves move along the size spectra from small to large body size as organisms grow.  Peaks of the waves are associated with slow growth (prey relatively rare) and low mortality (predators relatively rare).  As expected, the deterministic jump-growth time series gives a better match to the stochastic series than the McKendrick--von Foerster one, in terms of the period and shape of the waves (although they are not identical).

\subsection{Variable growth}\label{subsection.variable}

The jump-growth model and the McKendrick--von Foerster equation differ in that the former describes a process in which  organisms, starting at the same weight, develop different weights over the course of time.  In so doing, the jump-growth model captures an important feature of growth: when two organisms of the same weight eat prey items of different weights, the two organisms must subsequently have different weights. 

\begin{figure}[ht!]
	\centering
  	\includegraphics[width=5in, height=4in]{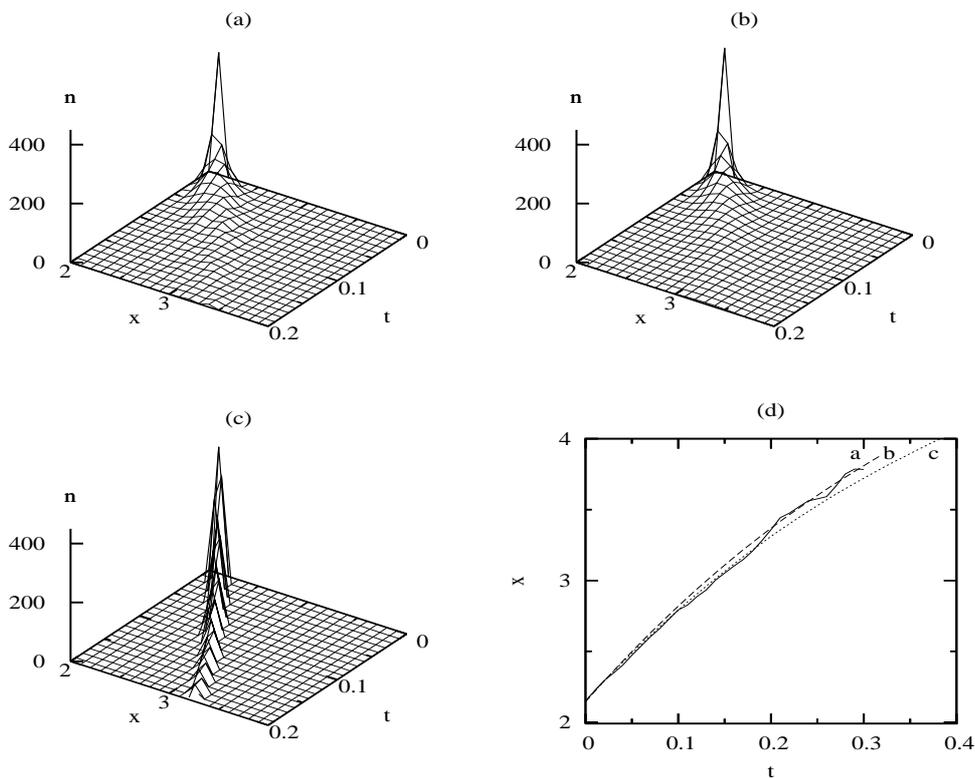}
		\caption{\small{
Number $n(x)$ of organisms with log weight $x$ over time $t$ in tagged cohorts embedded in size spectra.  Cohorts start in a weight range $2.1 \leq x < 2.2$ at $t=0$. (a) Stochastic jump-growth process; (b) deterministic jump-growth equation; (c) McKendrick--von Foerster equation; (d) mean weights over time computed for the cohorts shown in (a), (b), (c), and labelled correspondingly.  Parameter values given in Table \ref{table.parameters}.}}
  \label{fig.growth}
\end{figure}

Figure \ref{fig.growth} illustrates this feature of the models, using parameter values that highlight the differences between them. The results show the fate of a set of organisms that all start with very similar weights in the range $[x_d,x_d+\Delta')$;  the set can be thought of as a cohort which grows older without renewal.  In the stochastic jump-growth model, organisms were tagged individually, and the size distribution of the cohort over time was monitored.  In the deterministic jump-growth model we assumed a tagged cohort $u^*(x)$ at a density low enough relative to $u(x)$ for changes in $u^*(x)$ to come just from feeding on and being fed upon by $u(x)$, without any reciprocal effect of $u^*(x)$ on $u(x)$.  In the McKendrick--von Foerster simulation, differential equations for survival and growth in weight in the cohort were solved using the growth and death rates \eqref{mvfD} and \eqref{mvfG} respectively, as described in Law et al. \citeyearpar{Law:2008}.

The stochastic realisation (Figure \ref{fig.growth}a) shows the number of tagged individuals declining as time goes on (they are being eaten by larger organisms); it also shows the distribution of body weights spreading out.  The behaviour of the deterministic jump-growth equation matches the stochastic cohort closely (Figure \ref{fig.growth}b).  However, the McKendrick--von Foerster equation (Figure \ref{fig.growth}c) retains its initial spike-like distribution, because the growth trajectory from any size is fixed.

The average growth trajectories of all three models are close together (Figure \ref{fig.growth}d). As time goes on and the number of individuals in the stochastic cohort becomes small, fluctuations in the stochastic growth trajectory can be seen. Also, growth according to the McKendrick--von Foerster equation is slightly slower than in the deterministic jump-growth equation. However, these differences are small, and it is only when the second moments of growth are considered that the spreading in body sizes, missing from the McKendrick--von Foerster equation, becomes evident.

Adding the second-order diffusion term of (\ref{macro5}) to the McKendrick--von Foerster equation (\ref{mvf}) would recover the tendency for body size to spread.  However, this still leaves out higher order terms of the Taylor expansion (\ref{macro5}) which do not necessarily become small unless the steady state is an attractor.

\subsection{Dynamical stability}

\begin{figure}[htbp]
	\centering
 	\includegraphics[width=4in, height=5in]{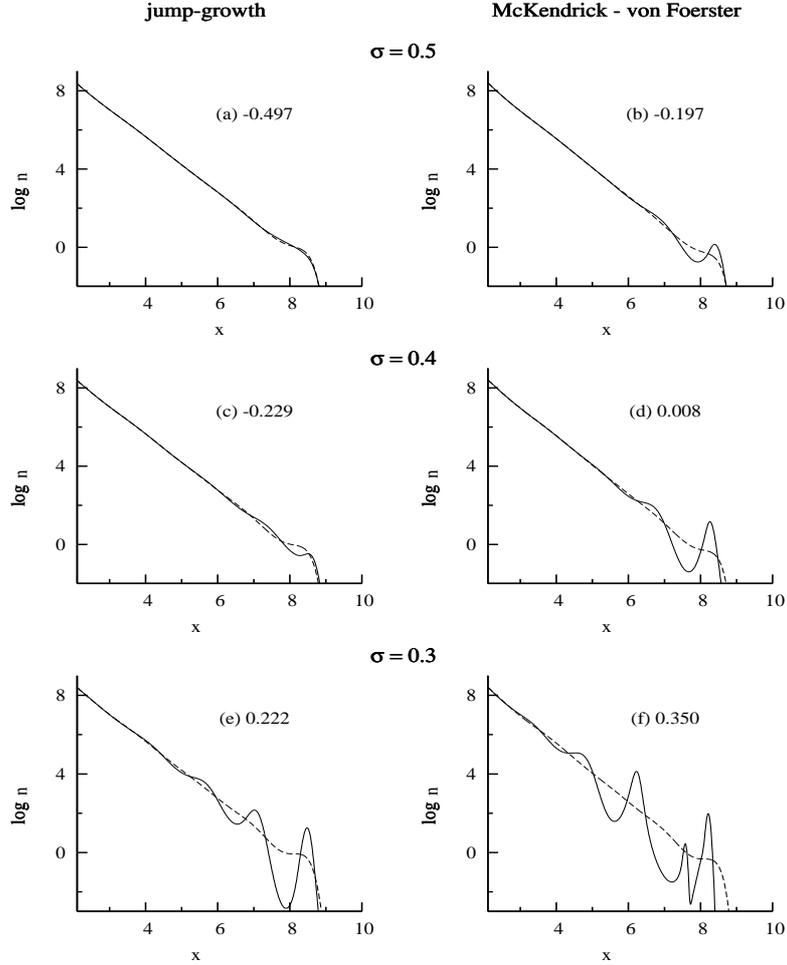}
		\caption{\small{
Steady-state size spectra (dashed lines), and transient size spectra (continuous lines) after a period of 5 time units has elapsed starting from the same initial function. Column 1 (a, c, e) obtained from the deterministic jump-growth equation; column 2 (b, d, f) obtained from the McKendrick--von Foerster equation. Diet breadths $\sigma$: 0.5 (a, b), 0.4 (c, d), 0.3 (e, f); other parameters given in Table \ref{table.parameters}. Steady states obtained by Newton-Raphson iteration, which also gives the Jacobian matrix at steady state \citep{Press:1992}; numbers given for each graph are $\max(Re(\lambda))$ where $\lambda$ is an eigenvalue of the Jacobian.}}
  \label{fig.stability}
\end{figure}

Figure \ref{fig.stability} gives examples of the steady states and stability properties of the jump-growth and McKendrick--von Foerster models. The breadth of diet $\sigma$ decreases from top to bottom in the figure.

At steady-state, the size spectra have similar shapes in the two models, and diet breadth has little effect on them.  For the most part the steady states are close to linear under the log transformation of both axes.  This linearity applies until near $x = 8$, where the extra size-dependent mortality starts to take effect.  In the region $2.1 \leq x < 7$ which is close to linear, the slopes are approximately $-1.42$ in the deterministic jump-growth equation and $-1.47$ in the McKendrick--von Foerster equation, equivalent to exponents $\gamma = 2.42$ and $\gamma = 2.47$ respectively.  These values are close to the value 2.47 predicted from analysis of the delta-function version of the feeding preference equation (\ref{gamma}).

Figure \ref{fig.stability} shows the existence of a bifurcation point at which the system flips from one dynamical regime to another as $\sigma$ changes.  For large enough $\sigma$ the steady state is an attractor, i.e. the Jacobian matrix evaluated at the steady state has $\max(Re(\lambda)) < 0$: size spectra initialised away from the steady state move towards it.  For small enough $\sigma$ this ceases to be the case, i.e. $\max(Re(\lambda)) > 0$; instead, the size spectra develop travelling waves like those seen in Figure \ref{fig.timeseries}.  Importantly, the bifurcation point occurs at a smaller value of $\sigma$ in the jump-growth equation. This may be because of the lack of spreading in body size in the McKendrick--von Foerster equation: such spreading would tend to dampen oscillations. A consequence of the difference is that a stability analysis of the McKendrick--von Foerster equation could be misleading;  see for instance Law et al. \citeyearpar{Law:2008}.  Although not shown here, the bifurcation to a travelling wave can also be induced by increasing the preferred ratio $B$ of the predator:prey body mass \citep{Law:2008}.

\section{Discussion}\label{section.discussion}

The starting point for our analysis was a simple, mechanistic, stochastic process in which a larger organism feeds on a smaller one, thereby causing the death of the prey and increment in its own weight.  From the master equation of the process, a macroscopic model for the dynamics of size spectra was derived, which we call the deterministic jump-growth model.  The equation is related to the Smoluchowski coagulation equation \citep{Smoluchowski:1916}, which describes how the size-distribution of inanimate coagulating particles changes over time.  However, the jump-growth equation has to deal with special features of living organisms, such as their ability to choose the size of their prey, and their inefficiency in turning these prey into their own body mass.  To cope with the vagaries of the animate world, the deterministic jump-growth equation is necessarily more general.

The expression for the steady-state derived from the deterministic jump-growth equation is consistent with the approximate constancy of biomass in logarithmic intervals of body mass often observed in marine ecosystems.  So the basic empirical regularity evidently follows from the bookkeeping of biomass, as it passes through the ecosystem.  However, the steady state may or may not be an attractor. As one might anticipate from the general oscillatory nature of predator-prey systems, another non-equilibrium attractor exists, here comprising waves of abundance that travel from small to large body size.  These waves have practical as well as theoretical interest in view of the large, often unexplained, fluctuations in exploited marine fish stocks (\citealp{Hsieh:2006, Anderson:2008, Blanchard:2008}, personal communication).

The jump-growth model is not the same as the McKendrick--von Foerster equation widely used in the study of dynamic size spectra.  This is because it allows organisms, starting at the same size, to become different through eating prey of different sizes.  The McKendrick--von Foerster equation, with its roots in age distributions \citep{McKendrick:1926, vonFoerster:1959} does not allow this:  organisms which start at the same age must always remain the same age.  An age-dependent McKendrick--von Foerster equation has been extended to allow for variable size at age \citep{Gurney:2007}, but this was by adding variability to a specified model of growth, the von Bertalanffy equation.  Growth of organisms in dynamic size spectra comes about in a quite different way, because it emerges directly from the action of predators feeding on prey.  This is not to suggest that variation in prey size is the only cause of variation in predator size;  in reality, a variety of extrinsic and intrinsic factors are most likely involved.

Although the deterministic jump-growth model is different from the  McKendrick--von Foerster equation, the latter can be derived from it using the lowest-order terms in a Taylor approximation.  The approximation requires that prey size is small relative to that of the predator, which will often apply in practice.  Thus for many purposes the McKendrick--von Foerster equation should work well, notwithstanding the numerical examples used in Section \ref{section.numerics} (deliberately chosen to contrast the two models).  This is with the caveat that higher-order terms in the Taylor expansion are not necessarily small when the attractor is a travelling-wave rather than a steady state, or when looking at spiky perturbations away from the steady state, even if prey are much smaller than their predators. To describe such non-equilibrium solutions accurately, the jump-growth model is needed.

When solving the jump-growth equation numerically, some care is needed in the discretisation of $\log w$.  Unlike the McKendrick--von Foerster equation, there is no guarantee that feeding will generate non-zero rate terms for growth.  If the multiplicative weight brackets $\Delta$ are too large relative to prey size, weight increments from feeding do not register, and an erroneous solution is obtained.  For a Gaussian feeding preference function (\ref{kernel}), a rule of thumb is that $\Delta$ needs to be of an order $K/(Be^{2\sigma})$ to capture properly the rate term due to growth of organisms.  Values of the order $B=10^2$, $\sigma = 0.5\log B$ and $K=0.1$ are realistic \citep{Paloheimo:1966, Cohen:1993, Jennings:2003}, requiring $\Delta$ to be of an order $10^{-5}$.  With marine size spectra encompassing ten orders of magnitude, numerical analyses clearly become demanding.  A small value of $B$ was used for the illustrations in Section \ref{section.numerics}, but it would be much harder to do the computations in a more realistic setting.

Faced with this difficulty, a halfway house would be to use the McKendrick--von Foerster equation with the diffusion term from the expansion in \eqref{macro5}. We are not aware of a previous derivation of the diffusion term for growth in body size, although diffusion in physical space has been considered in the context of the McKendrick--von Foerster equation \citep{Okubo:2001}.  Nor have we seen the use of a diffusion term in the McKendrick--von Foerster equation applied to size spectra, although the effects of introducing variability into Gompertz and von Bertalanffy growth models have been described \citep{Bardos:2005, Gurney:2007}.  It would be instructive to know how much the McKendrick--von Foerster approximation could be improved by introducing this extra term.

Several further features of real-world ecosystems, not dealt with in this paper, will modify our results.  First, some feedback between the abundance of phytoplankton and consumers is to be expected.  Second, perfect compensation in renewal of consumers is unlikely, especially when travelling waves affect the abundance of reproducing individuals.  Such processes generate long, potentially destabilizing, feedback loops.  Third, consumers do not all start life with the same potential for growth and reproduction.  They comprise a number of different species with different life histories \citep{Andersen:2006, Blanchard:2008}. They are born at different sizes, they grow to different sizes, and they allocate different proportions of their limited resources to growth, maintenance and reproduction along the way \citep{ Maury:2007}.  Such processes loosen the dynamical coupling between a feeding organism and its prey.  

There is much to learn about the intricacies of biology that can stabilize and destabilize  marine ecosystems.  It is important to obtain this knowledge because the biomass in such ecosystems is typically of major economic importance, heavily exploited, and with dynamics that are not well understood.  The deterministic jump-growth equation derived here should place this programme of research on a more rigorous footing.

\addvspace{0.2 in}
\noindent \textbf{Acknowledgements}: We thank Julia Blanchard, Jennifer Burrow, Alex James, Jon Pitchford and Michael Plank and the referees of the paper for many helpful insights, and Kai Wirtz for pointing out the relation to the Smoluchowski coagulation equation.  The research was supported by a studentship to SD from the Natural Environment Research Council UK, with the Centre for Environment Fisheries and Aquaculture Science UK as the CASE partner.  RL was supported by the Royal Society of New Zealand Marsden Fund, grant 08-UOC-034.

\appendix
\section*{Appendix A: Derivation of Langevin equation}\label{gillespie}

Our treatment of the jump-growth model using the van Kampen expansion in Section \ref{subsection.sep} did not provide a justification for assuming that the fluctuations $\xib$ around the solution $\phib$ of the deterministic equation \eqref{macro2} are damped by a factor of $\Omega^{1/2}$. In this appendix we derive an approximate stochastic differential equation for the jump-growth model, adapting an approximation procedure used by Gillespie \citeyearpar{Gillespie:2000} for stochastic models of chemical reactions. We will find that the deterministic part of the equation coincides with our deterministic jump-growth equation \eqref{macro2} and that the stochastic noise term is indeed suppressed by a factor of $\Omega^{1/2}$.

Because of the stochastic nature of the jump-growth model, the vector of numbers "$[\dots,n_{-1},n_0,n_1,\dots]$ in each weight bracket introduced in subsection (\ref{subsection.master}) is described by a stochastic process $\nv(t)$. In a time interval $[t,t+\tau]$ a number of predation events will take place, each of which changes the numbers. This is expressed by the equation
\begin{equation}\label{n1}
 \nv(t+\tau) = \nv(t) + \sum_{i,j}R_{ij}(\nv(t),\tau)\nuv_{ij},
\end{equation}
where the $R_{ij}(\nv,\tau)$ are random variables giving the number of predation events taking place in the time interval $[t,t+\tau]$ that involve a predator from weight bracket $i$ and a prey from weight bracket $j$. The $\nuv_{ij}$ are the vectors that give the change in numbers caused by such a predation process, as described in Subsection (\ref{subsection.master}).  We now will argue that the random variables $R_{ij}(\nv(t),\tau)$ can be approximated by normally distributed variables.

The rate  $a_{ij}$ of each individual predation event depends on the numbers of individuals
\begin{equation}\label{prop}
 a_{ij}(\nv) = \Omega^{-1}k_{ij} n_i n_j.
\end{equation} 
As the numbers change after each event, the events are unfortunately not independent. However, because the numbers change only by $\pm 1$ in each event, the change to the rates is very small if the numbers are large. So, if we choose the time span $\tau$ small enough so that not too many predation events take place, the rates can be approximated as remaining constant throughout the time interval,
\begin{equation}
 a_{ij}(\nv(t')) \approx a_{ij}(\nv(t)) ~~~\forall t'\in[t,t+\tau].
\end{equation}
In that case the predation events can be treated as independent and therefore the number $R_{ij}(\nv(t),\tau)$ of event taking place in the time interval follows the Poisson distribution with parameter $\tau a_{ij}(\nv(t))$.

Next we assume that the parameter $\tau a_{ij}(\nv(t))$ is either zero or large enough so that the Poisson distribution is well approximated by the normal distribution with mean and variance both equal to $\tau a_{ij}(\nv(t))$. Again this is easy to justify when the numbers are large and provided the feeding kernel $k_{ij}$ is bounded away from zero. In our case, where the feeding kernel contains a Gaussian, we need to neglect the rare events in the tails of the Gaussian.

Note that we are placing two opposing conditions on the size of the time interval $\tau$: it needs to be both small enough so that the rates don't change much but also large enough so that the number of events can be taken to be normally distributed. Such an intermediate range for $\tau$ will exist, provided the numbers of individuals in the weight brackets are large enough. In our application, where the overall number of individuals involved is truly huge, our approximations will be very good except for very large weights where the density is very small and where the approximations will break down.

Now that we have argued that the $R_{ij}$ are well approximated by normal random variables with mean and variance both equal to $\tau a_{ij}(\nv(t))$, we express them as
\begin{equation}
 R_{ij}(\nv(t),\tau) =  a_{ij}(\nv(t))\tau +\sqrt{a_{ij}(\nv(t))\tau} \ r_{ij}
\end{equation}
where the $r_{ij}$ are normal random variables with mean $0$ and variance $1$. Substituting this into \eqref{n1}, rearranging terms and dividing by $\tau$ gives
\begin{equation}
 \frac{\nv(t+\tau)-\nv(t)}{\tau} = \sum_{ij}a_{ij}(\nv(t))\nuv_{ij}+
  \sum_{ij}\sqrt{a_{ij}(\nv(t))}\nuv_{ij}\tau^{-1/2}r_{ij}.
\end{equation}
We now approximate this equation, which is valid for small but finite $\tau$, by the stochastic differential equation obtained by taking the limit $\tau\rightarrow0$,
\begin{equation}\label{langevin}
 \frac{d\nv(t)}{dt} = \sum_{ij}a_{ij}(\nv(t))\nuv_{ij}+
  \sum_{ij}\sqrt{a_{ij}(\nv(t))}\nuv_{ij}\eta_{ij}(t),
\end{equation}
where $\eta_{ij}(t)$ are independent white noise processes. This type of equation is known as a Langevin equation, see for example \cite{vanKampen:1992}.

Substituting the explicit expressions \eqref{prop} for the rates into the Langevin equation \eqref{langevin} gives
\begin{align}
 \frac{dn_i}{dt} = &\Omega^{-1}\sum_{j}\left(-k_{ij}n_in_j-k_{ji}n_jn_i
  +k_{mj}n_mn_j\right)\\
  &+\Omega^{-1/2}\sum_j\left(-\sqrt{k_{ij}n_in_j}\eta_{ij}
  -\sqrt{k_{ji}n_jn_i}\eta_{ji}+\sqrt{k_{mj}n_mn_j}\eta_{mj}\right).\nonumber
\end{align}
When we write the equation in terms of the population densities $\Phi_i=\Omega^{-1}n_i$ we see that the fluctuation terms are suppressed by a factor of $\Omega^{-1/2}$.
\begin{align*}
 \frac{d\Phi_i}{dt} = &\sum_{j}\left(-k_{ij}\Phi_i\Phi_j-k_{ji}\Phi_j\Phi_i
  +k_{mj}\Phi_m\Phi_j\right)\\
  &+\Omega^{-1/2}\sum_j\left(-\sqrt{k_{ij}\Phi_i\Phi_j}\eta_{ij}
  -\sqrt{k_{ji}\Phi_j\Phi_i}\eta_{ji}+\sqrt{k_{mj}\Phi_m\Phi_j}\eta_{mj}\right).
\end{align*}
For large system size $\Omega$ the fluctuation terms can be neglected and we end up with our equation \eqref{macro2}.

\addcontentsline{toc}{chapter}{\numberline{}Bibliography}
\bibliographystyle{elsart-harv} 
\bibliography{bibfile}

\end{document}